 \documentclass[letterpaper, 10 pt, conference]{ieeeconf}
\IEEEoverridecommandlockouts               % This command is only needed if
\usepackage{setspace}
\usepackage{cite}
\usepackage{amsmath,amssymb,amsfonts}
\usepackage{algorithmic}
\usepackage{graphicx}
\usepackage{subfigure}
\usepackage{textcomp}
\usepackage{bm}
\usepackage{enumerate}
\usepackage{color}

\def\BibTeX{{\rm B\kern-.05em{\sc i\kern-.025em b}\kern-.08em
    T\kern-.1667em\lower.7ex\hbox{E}\kern-.125emX}}

\usepackage[linesnumbered,ruled]{algorithm2e}
\usepackage{mathrsfs}
\usepackage{booktabs}
\newtheorem{theorem}{\textbf{Theorem}}
\newtheorem{assumption}{\textbf{Assumption}}

\newtheorem{remark}{\textbf{Remark}}

\def\BibTeX{{\rm B\kern-.05em{\sc i\kern-.025em b}\kern-.08em
    T\kern-.1667em\lower.7ex\hbox{E}\kern-.125emX}}
\begin{document}
\title{\LARGE \bf
% Topology Inference for Multi-agent Cooperation under \\ Unmeasurable Latent Input\\
% Topology Inference for Cooperative Control with \\Unmeasurable Latent Input
Topology Inference for Network Systems with Unknown Inputs
}
\author{Qing Jiao$^\dagger$, Yushan Li$^\ddagger$, and Jianping He$^\dagger$
    \thanks{This work was supported by the National Natural Science Foundation of China under Grant 62373247. The work of Yushan Li was also supported by the Outstanding Ph.D. Graduate Development Scholarship from Shanghai Jiao Tong University. \textit{(Corresponding author: Yushan Li.)}}
	\thanks{$^\dagger$: Qing Jiao and Jianping He are with the Department of Automation, Shanghai Jiao Tong University, and Key Laboratory of System Control and Information Processing, Ministry of Education of China, Shanghai, China. E-mail: \{jiaoqing, jphe\}@sjtu.edu.cn.}
    \thanks{$^\ddagger$: Yushan Li is with the Division of Decision and Control Systems, KTH Royal Institute of Technology, Sweden. E-mail: yushanl@kth.se.}
    % \thanks{Preliminary results have been accepted by IEEE VTC 2021-Fall \cite{qingjiao2021topology}.}
}%

\maketitle
\thispagestyle{empty}
\pagestyle{empty}
\begin{abstract}
Topology inference is a powerful tool to better understand the behaviours of network systems (NSs). 
Different from most of prior works, 
this paper is dedicated to inferring the directed topology of NSs from noisy observations, where the nodes are influenced by unknown time-varying inputs. 
These inputs can be actively injected signals by the user, intrinsic system noises or extrinsic environment interference. 
To tackle this challenging problem, we propose a two-stage inference scheme to overcome the influence of the inputs. 
First, by leveraging the second-order difference of the state evolution, we establish a judging criterion to detect the input injection time and provide the probability guarantees. 
With this injection time to determine available observations, an initial topology is accordingly inferred to further facilitate the input estimation. 
Second, utilizing the stability characteristic of the system response, 
a recursive input filtering algorithm is designed to approximate the zero-input response, which directly reflects the topology structure. 
Then, we construct a decreasing-weight based optimization problem to infer the final network topology from the approximated response. 
Comprehensive simulations demonstrate the effectiveness of the proposed method. 
\end{abstract}

\section{Introduction}

% \textcolor{blue}{
The topology of network systems (NSs) characterizes the key interaction structure among nodes and greatly influence the network dynamics. 
In recent years, inferring the network topology has become a crucial cornerstone and fundamental step for a deeper understanding of NSs, and received considerable attention across many real-world networks \cite{cui2024topology}. 
% }
% For networked systems (NSs), the underlying topology characterizes the key interaction information among network nodes and greatly influence the network behaviors.   
% In recent years, inferring the network topology has received considerable attention across many real-world networks \cite{cui2024topology}, which a crucial cornerstone and fundamental step for a deeper understanding of NSs. 
For example, it can be used to identify the hidden buses in electric distribution grids \cite{anguluri2021grid}, detect the influence of users in social networks \cite{ogburn2024causal}, and identify the active links in sensor networks \cite{du2023network}. 
% Usually, the network topology of NSs is not directly available \cite{van2018identifiability}, thus, inferring the network topology is a crucial cornerstone and fundamental step for a deeper understanding of NSs. 
% e.g., electric distribution grids \cite{anguluri2021grid}, 
% \cite{anguluri2021grid,cavraro2018graph}, 
% social networks \cite{ogburn2024causal}, and sensor networks \cite{du2023network}.  
% and brain networks \cite{vosoughi2020large}. 
% \textcolor{blue}{
In this work, 
we consider that the NS is involved with unknown time-varying inputs, 
and aim to provide an effective topology inference method that overcomes the influence of the unknown inputs.
% and aim to provide reliable topology inference design that can overcome the influence of the unknown inputs. 
% }
% In this work, 
% we consider inferring the network topology when unknown time-varying inputs are injected into the NSs, 
% and provide reliable topology inference by filtering the unknown inputs from observations.

% \subsection{Related Works} 
Various works that infer the network topology from the observations of NSs have been developed in the literature. 
A common way is to utilize specific signals to excite the NS, 
where the statistical characteristics of these signals are usually required to be known. 
% When utilizing specific signals to realize the topology inference, 
% the statistical characteristics of these signals are usually required to be known.   
For example, the information about the external input \cite{li2023local} and intrinsic noise \cite{hayden2016network} are assumed to be available.  
% initial agent state \cite{Santiago2018Network},
The transfer entropy (TE), proposed by \cite{schreiber2000measuring}, provides a computationally-inexpensive way for causal network inference based on the information-theoretic measure, 
and many TE-based topology inference approaches have been proposed \cite{sharma2019communication,h2020topological}.   
% is one of the most prominent approaches to capture the casual relationships between network nodes
Granger causality is another well-known tool for causal inference \cite{dimovska2020control},
% \cite{liu2022topology,dimovska2020control},  
which take the task of topology inference as learning temporal causal structures among multiple time series \cite{testi2020blind}. 
% There are extensive research on topology inference for various dynamical models.
Besides, graph signal processing (GSP) approaches offer a new perspective to infer topology from observations \cite{Sandryhaila2013Discrete}, 
which model the observations as graph signals and the dynamic processes as the description of physical systems \cite{Zhu2020Network}.
% Besides, graph signal processing (GSP) approaches offer a new perspective on the problem of topology inference from the observed signals \cite{Sandryhaila2013Discrete}.
% The approaches leveraging GSP tools usually model observations as graph signals and model dynamic process as the description of a physical system.
% Topology inference utilize  GSP mainly based on three aspects \cite{Zhu2020Network}:
% signal smoothness \cite{Dong2016Learning},
% causal dependency \cite{Pasdeloup2018Characterization,Egilmez2018Graph},
% and network diffusion \cite{Thanou2017Learning,ShafipourNetwork}.
Among these GSP-based methods, assumptions on the structure of network and the properties of signals play an important role in determining the performance of the inference results, e.g., the sparsity of networks \cite{Chepuri2017Learning} and the smoothness of signals \cite{dong2016learning}. 
Other prominent approaches such as Bayesian approach \cite{liu2023topology}, spectral analysis \cite{su2021identification} and the sparse maximum likelihood estimation \cite{Matsuzaki2024} provide significant ideas to infer the network topology for NSs.

Despite the prominent contributions of these pioneering works, most of them have not considered the situations where the NS is influenced by external inputs. 
These inputs could be the state-independent inputs to achieve preset goals (e.g., using a reference dynamics as the virtual leader in a multi-agent system \cite{movric2014cooperative}), random noises and disturbances (e.g., random noises in communication process among the nodes \cite{Zhu2020Network}), 
or designed injected signals (e.g., active excitation inputs to produce new system responses \cite{restrepo2023simultaneous}). 
Several works have investigated topology inference for NSs with inputs involved, where sufficient observations are available and some prior information of the inputs are assumed known, 
% focused on NSs with inputs involved and aim to infer the topology from sufficient observations, with the assumption that some prior information of the inputs are known, 
e.g., the inputs are constant \cite{li2023local}, or the statistical characteristic of the random inputs are unknown \cite{li2024topology}. 
However, if the inputs are totally unknown and time-varying, it will hinder the ability of aforementioned works to obtain reliable topology estimates. 
Hence, how to infer the topology of NSs subject to unknown inputs from limited observations is still an open problem.

Motivated by the above issues, this paper focuses on inferring the topology of NSs with time-varying inputs while without any prior information of the inputs. 
% investigates the topology inference for directed NSs with time-varying inputs, while without prior information of the inputs. 
Our key insight is to develop a two-stage inference scheme: estimate the unknown inputs, and utilize the observation with inputs filtered to obtain a final reliable topology.
The contributions are summarized as follows. 
First, we characterize the state evolution of the NS into two parts: zero-input and zero-state responses. 
Based on the stability characteristic of the zero-input response, a judging criterion with probability guarantees is established to detect the input injection time. 
Second, we leverage the inferred injection time to obtain an initial topology, which facilitates estimating the inputs in the dynamic process. 
A recursive input filtering algorithm is further designed to approximate the zero-input response. 
Finally, we propose a decreasing-weight based optimization design to infer the topology from the approximated response. 
Extensive and comparative simulations verify the effectiveness of the proposed method under various unknown inputs. 

The remainder of this paper is as follows. 
In Section \ref{sec:Preliminaries}, some preliminaries of NSs and problem modeling are presented. 
The inference method and theoretical analysis are provided in Section \ref{sec:Design}.
Simulation results are shown in Section \ref{sec:simulation}.
Finally, Section \ref{sec:Conclusion} concludes the paper.

\section{Preliminaries and Problem Formulation} \label{sec:Preliminaries}
\subsection{System Model}
% There are three necessary preliminaries.
% The basic ideas and some main results are described as follows.

Let $\mathcal{G}(\mathcal{V},\mathcal{E})$ be a directed graph which models the network system, where $\mathcal{V}=\{1, \ldots, n\}$ is the set of $n$ nodes and  $\mathcal{E}\subseteq \mathcal{V}\times\mathcal{V}$ is the set of edges.
The edge $(i,j)\in \mathcal{E}$ is associated with the weight ${w}_{ij}$, and ${w}_{ij}\!\ne\! 0$ indicates that node $i$ receives information from node $j$. 
The set $\mathcal{N}_i=\{j|{w}_{ij} \ne 0,j\in\mathcal{V}\}$ consists of all in-neighbors of node $i$, 
and the topology matrix $W=[w_{ij}]_{i,j=1}^n$ is a row-stochastic matrix. 
$\mathbf { 1}$ and $I$ refer to all-one column vector and identity matrix, respectively.

Then, the state evolution of node $i$ is described by 
\begin{equation}\label{eq:model}
x_{t+1}^i  = \sum\nolimits_{j \in \mathcal{N}_i } w_{ij}{x_t^j}+u^i_t,
\end{equation}
where $x_t^i \in \mathbb{R} $ is the state of node $i$ at time $t$, and $u_t^i \in \mathbb{R}$ ($|u_t^i|<\infty$) is the unknown input injected on node $i$. 
This type of model is common to see, e.g., in leader-follower formation control \cite{movric2014cooperative} and secure consensus problem when extra disturbances are involved \cite{Zhu2020Network}.  
Accordingly, the global form of the system dynamics is given by
\begin{align}\label{eq:separation}
x_{t}&=Wx_{t-1}+u_{t-1} \nonumber \\
&= \underbrace{W^t x_0}_{ x_t^* } + \underbrace{  \sum\nolimits_{\ell = 0}^{t-1} W^{t-1-\ell} u_{\ell}  }_{ x_t^u },
\end{align}
where $x_t^*$ and $x_t^u$ can be seen as the zero-input and the zero-state responses of the system, respectively. 
% Notice that if $u_t=0$ for all $t$, then the system state (denoted as $x_t^*$) will converge to the following constant vector
% \begin{equation}
% \lim_{t\to\infty} x_t^*= \lim_{t\to\infty}  W^t x_0 =  (p^\mathsf{T} x_0) \bm{1},
% \end{equation}
% where $ p\in \mathbb{R}^n $ is the left dominant eigenvector of $W$. 
The following assumptions are made throughout this paper. 
\begin{assumption}\label{assu:topo}
The topology matrix $W$ is row-stochastic and has a simple eigenvalue $1$. 
% For the sequence $\{u_t\}_{t=0}^{T}$, there exists $s$ ($1 \le s \le T$) moments such that the inputs at these moments are zero. 
% The input sequence satisfies 
\end{assumption}
\begin{assumption}\label{assu:input}
% The topology matrix $W$ is row-stochastic and has a simple eigenvalue $1$. 
The nodal input sequence $\{u_t^i\}_{t=0}^{T-1}$ is Lipschitz continuous on $t\in\mathbb{R}$, and it is injected after time $s$ ($s >1$). 
% The input sequence satisfies 
\end{assumption}

Note that Assumption \ref{assu:topo} is a common topology setting in NSs \cite{FB-LNS}, and it follows that the zero-input response $x_t^*$ will converge to a constant vector, $\lim_{t\to\infty} x_t^*= \lim_{t\to\infty}  W^t x_0 =  (p^\mathsf{T} x_0) \bm{1}$, 
% \begin{equation}
% \lim_{t\to\infty} x_{t,*}= \lim_{t\to\infty}  W^t x_0 =  (p^\mathsf{T} x_0) \bm{1},
% \end{equation}
where $ p\in \mathbb{R}^n $ is the left dominant eigenvector of $W$. 
Assumption \ref{assu:input} is a mild condition to accommodate various kinds of inputs. 

% avoid that all the time-varying inputs are non-zero. 
% This situation is easy to find, e.g., the inputs are not necessarily injected at the initial moment, or the input sequence subjects to certain functions that will equal to zero at some moments. 

\subsection{Problem of Interest}

Consider that an external observer can access the system states with observation noises, given by
\begin{equation}
y_t= x_t + \varepsilon_t,
\end{equation}
where  $\varepsilon_t$ is the independent and identically distributed Gaussian noise, satisfying $ \varepsilon_t \sim N(0,\sigma ^2I)$. 
% For every two consecutive observations, it holds that 
% \begin{align}
% y_t &= x_t + \varepsilon_t=W(y_{t-1}-\varepsilon_{t-1}) + u_{t-1} \varepsilon_t \nonumber \\
% &= Wy_{t-1} + \omega_t,
% \end{align}
For every two consecutive observations, it holds that
\begin{align} \label{eq:expand_form}
y_{t+1} &=(W x_{t} + u_t) +\varepsilon_{t+1} =W (y_{t}-\varepsilon_{t})+ u_t + \varepsilon_{t+1} \nonumber \\
& = W y_{t}+ u_t+ \varepsilon_{t+1}-W\varepsilon_{t}, 
\end{align}
where $\varepsilon_{t+1}-W\varepsilon_{t}\sim N(0,( \sigma^2 I+WW^\mathsf{T}\sigma^2 ))$ can be regarded as a composite observation noise vector. 
Note that \eqref{eq:expand_form} only describes the quantitative relationship of two consecutive observations, and does not represent a dynamic process. 

Finally, the goal of this paper is to infer the topology matrix $W$ from the collected observations $\{y_t\}_{t=0}^{T}$, where the input set $\mathcal{U}=\{u_t\}_{t=0}^{T-1}$ is unknown. 
Mathematically, we utilize the formulation \eqref{eq:expand_form} and turn to solve the following optimization problem 
\begin{subequations}\label{eq:prob}
\begin{align}
{\Phi}_{0}:~\mathop {\min }\limits_{ W, \mathcal{U} } ~& \sum\limits_{t = 0}^{T-1} \left\| y_{t+1}- W y_{t}- u_t \right\|_2^2  \! +\! \rho \sum\limits_{i = 1}^{n}{\left\| W_{[i]} \right\|_1}\label{eq:prob_a} \\
\text{s.t.}~&W \mathbf { 1} =\mathbf{ 1}, \label{eq:prob_b}
\end{align}
\end{subequations}
where $W_{[i]}$ represents the $i$-th row of $W$, the term $\sum\limits_{t = 0}^{T-1} \left\| y_{t+1}- W y_{t}- u_t \right\|_2^2$ is the empirical risk, and $\rho \sum\limits_{i = 1}^{n}{\left\| W_{[i]} \right\|_1}$ is the regularization term characterizing topology sparsity ($\rho>0$ is the regularization parameter). 
% , the two sum terms in \eqref{eq:prob_a} represent the empirical risk and the regularization term (characterizing specific topology properties), respectively, and $\rho>0$ is the regularization parameter. 
The constraint \eqref{eq:prob_b} ensures the row-stochasticity of $W$.

The considered problem ${\Phi}_{0}$ is challenging because the time-varying input is totally unknown to the observer and will also accumulate in the state evolution, making it extremely hard to infer the topology. 
To address these issues, our key idea is to leverage the row-stochasticity of $W$ to construct an initial topology estimate, and then design a zero-input response based inference algorithm to infer the underlying topology. 
Note that if the input is time-invariant, the situation will be much easier (e.g., the work \cite{li2023local} has well addressed this problem), and there is no need to use the method in this paper. 
\section{Inference Method Design} \label{sec:Design}
This section presents the two-stage inference scheme. 
First, a judging criterion is designed to detect the input injection,
which is used to obtain an initial topology estimate. 
Second, the initial estimate is leveraged to approximate the zero-input response, which supports to obtain a reliable final topology.

% In this section, 
% we first design a judging criterion to detect the input injection, and use the determined injection time to obtain an initial topology estimate. 
% Then, we leverage the initial estimate to approximate the zero-input response, and infer the final topology effectively. 

\subsection{Initial Topology Inference}

% Recall that two consecutive observations satisfy \eqref{eq:expand_form}. 
% Let $y_{t+1}^{*,i}= [W y_{t}]^i + \omega_{t}^i$ be node $i$'s observation corresponding to the situation $u_t^i=0$, where the variance of $\omega_{t}^i$ is upper bounded by 
% \begin{align}\label{eq:noise_bound}
% (\sigma_{\omega}^i)^2=\left(1\!+\!\sum\limits_{\ell  = 1}^{n} W_{i\ell}^2 \right)\sigma^2 \le 2 \sigma^2 = \bar{\sigma}^2. 
% \end{align}

When the inputs $\{u_t\}_{t=0}^{T-1}$ are totally unknown, finding accurate solutions to ${\Phi}_{0}$ would be challenging. 
Since the inputs are considered absent at the initial time, there exists an input injection time during the dynamic process. 
This point can be exploited to obtain a rough initial topology without knowing the inputs, and then we use this initial estimate to further improve the topology inference performance. 

Notice that the input in model \eqref{eq:model} reflects the first-order information of the NS. 
Hence, we begin with using the first-order observation difference of node $i$ to facilitate the method design, given by 
\begin{equation}
\Delta y_{t}^{i} = y_{t+1}^{i}-y_{t}^{i},~t\in\mathbb{N}. 
\end{equation}
Then, the following theorem quantifies the deviation of two groups of observation differences when the inputs are not injected yet.   
% quantify  the second-order observation difference when the inputs are not injected yet. 
% to quantify  the second-order observation difference 
\begin{theorem}\label{th:input_time}
% Suppose the inputs are zeros before time $t+1$,
If the inputs $u_{t'}=0, t'=0,\cdots,t$, 
then the observation difference satisfies the following property
\begin{align}\label{eq:th1}
\Pr \{ |\Delta y_{t}^{i} - \Delta y_{t-1}^{i} |< \Delta^2 y_{t -1}^{\max} +  c_{\epsilon} \bar{\sigma} \} \ge 1-\frac{1}{c_{\epsilon}^2}, 
\end{align}
where $\Delta^2 y_{t -1}^{\max}\!= \!\max\{ | \Delta y_{t-1}^{\ell } \!-\! \Delta y_{t-1}^{j}|\!: j,\ell \!\in\! \mathcal{V}\} $, $c_{\epsilon}>0$ is a constant, and $\bar{\sigma} \! =\! \sqrt{7} \sigma $. 
\end{theorem}

\begin{proof}
First, given two groups of observations $\Delta y_{t}^{i}$ and $\Delta y_{t-1}^{i}$, it holds that 
\begin{align}
\Delta y_{t}^{i} &= [W \Delta y_{t-1}]^{i} + \varepsilon_{t+1}^i -\varepsilon_{t}^i -[W(\varepsilon_{t}-\varepsilon_{t-1})]^i   + u_t^i - u_{t-1}^i \nonumber \\
&= [W \Delta y_{t-1}]^{i} + \omega_{t}^i + u_t^i - u_{t-1}^i,
\end{align}
where $\omega_{t}^i= \varepsilon_{t+1}^i -[(W+I)\varepsilon_{t}]^i+[W\varepsilon_{t-1}]^i $ is a composite noise variable. 
Notice that $\omega_{t}^i$ is still a zero-mean Gaussian noise and its variance satisfies 
\begin{align}\label{eq:noise_bound0}
(\sigma_{\omega}^i)^2 & =\left(1\!+  (W_{ii}+1)^2 + \!\sum\limits_{\ell = 1,\ell\neq i}^{n} W_{i\ell}^2 +  \sum\limits_{\ell = 1}^{n} W_{i\ell}^2 \right)\sigma^2   \nonumber \\
&\le  (1+4+1+1) \sigma^2 =7 \sigma^2 = \bar{\sigma}^2. 
\end{align}
Since the inputs are zeros before time $t+1$, we have 
\begin{align}\label{eq:deviation_dif}
\Delta y_{t}^{i} - \Delta y_{t-1}^{i}  &= [W \Delta y_{t-1} ]^{i}- \Delta y_{t-1}^{i} +  \omega_{t}^{i}  \nonumber \\
 \Rightarrow   |\Delta y_{t}^{i} - \Delta y_{t-1}^{i} | &\le |[W \Delta y_{t-1} ]^{i}- \Delta y_{t-1}^{i}| + |\omega_{t}^{i}| \nonumber \\
 & \le \Delta^2{y}_{t-1}^{\max}   + |\omega_{t}^{i}|,
 % \omega_{t}^{i} -\Delta^2{y}_{t}^{\max}    \le   \Delta y_{t}^{i} - \Delta y_{t-1}^{i} \le \omega_{t}^{i}  + \Delta \tilde{y}_{t}^{\max} .     
\end{align}
where the last inequality utilities the row-stochasticity of $W$. 
% where $\Delta \tilde{y}_{t-1}^{\max}=  \max\{ | \Delta y_{t-1}^{\ell }- \Delta y_{t-1}^{j}|: j,\ell \in \mathcal{V}\} $. 
Applying the famous Chebyshev inequality on $\omega_{t}^{i}$ yields that 
\begin{align}\label{eq:noise_bound}
\Pr \{ |\omega_{t}^{i}|<\epsilon \}\ge 1-\frac{(\sigma_{\omega}^i)^2 }{\epsilon^2},
\end{align}
where $\epsilon>0$ is a constant. 
Setting $\epsilon=c_{\epsilon} \sigma_{\omega}^i$ and utilizing \eqref{eq:noise_bound0},
% substituting it into \eqref{eq:noise_bound}, 
then it follows that 
\begin{align}\label{eq:no_input}
\Pr \{ |\omega_{t}^{i}|<c_{\epsilon} \bar{\sigma} \} &=  \Pr \{ \Delta^2{y}_{t-1}^{\max}  +  |\omega_{t}^{i}|< \Delta^2{y}_{t-1}^{\max}  +c_{\epsilon} \bar{\sigma} \} \nonumber \\
&\ge \Pr \{ |\omega_{t}^{i}|<c_{\epsilon} \sigma_{\omega}^i \} \ge 1-\frac{1}{c_{\epsilon}^2}.  
\end{align}
% where the first row and second row utilize \eqref{eq:deviation_dif} and \eqref{eq:noise_bound0}, respectively. 
Finally, 
% since $|\Delta y_{t}^{i} - \Delta y_{t-1}^{i} | \le \Delta^2{y}_{t}^{\max}   + |\omega_{t}^{i}|$ strictly holds, 
one can directly obtain \eqref{eq:th1} from \eqref{eq:deviation_dif} and \eqref{eq:no_input}. 
The proof is completed.
% \begin{align}\label{eq:no_input}
% \Pr\{|\Delta y_{t}^{i} - \Delta y_{t-1}^{i} | < \Delta^2{y}_{t}^{\max}  +c_{\epsilon} \bar{\sigma} \}&
% \Pr \{ |\omega_{t}^{i}|<c_{\epsilon} \bar{\sigma} \} =  \Pr \{ \Delta^2{y}_{t}^{\max}  +  |\omega_{t}^{i}|< \Delta^2{y}_{t}^{\max}  +c_{\epsilon} \bar{\sigma} \} \nonumber \\
% =&\ge \Pr \{ |\omega_{t}^{i}|<c_{\epsilon} \sigma_{\omega}^i \} \ge 1-\frac{1}{c_{\epsilon}^2} ,
% % \Rightarrow &  =\Pr \{ |\omega_{t}^{i}|<c_{\epsilon} \bar{\sigma} \} 
% % \Pr \{ |\Delta y_{t}^{i} - \Delta y_{t-1}^{i} |< \Delta \tilde{y}_{t-1}^{\max} +  c_{\epsilon} \bar{\sigma} \} \ge 1-\frac{1}{c_{\epsilon}^2}.
% \end{align}
\end{proof}

Based on Theorem \ref{th:input_time}, it is straight to obtain that if $\{u_{t'},t'=1,\cdots,t-1\}$ are zeros, then \eqref{eq:no_input} holds. 
Notice that this conclusion also works for other zero-mean non-Gaussian noises. 
For example, taking $c_{\epsilon}=3$, we have 
\begin{equation}\label{eq:pr_bound}
\Pr \{ |\Delta y_{t}^{i} - \Delta y_{t-1}^{i} |<  s_t \} \ge 8/9,
\end{equation}
where $s_t=\Delta^2 {y}_{t-1}^{\max} +  3 \bar{\sigma}$.
% $\Pr \{ |\Delta y_{t}^{i} - \Delta y_{t-1}^{i} |< \Delta \tilde{y}_{t-1}^{\max} +  3 \bar{\sigma} \} \ge 8/9$. 
Specifically, when the noise is Gaussian, the probability bound in \eqref{eq:pr_bound} can be further increased to $0.997$ according to the well-known $3\sigma$-rule \cite{pukelsheim1994three}. 
For simplicity, we directly take $s_t$ as the maximum bound of $|\Delta y_{t}^{i} - \Delta y_{t-1}^{i} |$, which is empirically treated as near certainty in practical experiments. 
By leveraging the converse-negative version of the above property, 
if $|\Delta y_{t}^{i} - \Delta y_{t-1}^{i} |<  s_t$ is not satisfied at time $t$, then it can be inferred that $u_t^i\neq0$.
Motivated by this point, the following criterion rule is designed to detect the injecting time of the input on node $i$
\begin{equation}\label{eq:input_time}
\tau_i=\inf\{t:  |\Delta y_{t}^{i} - \Delta y_{t-1}^{i} | \ge s_t, t\in\mathbb{N}^+\}.
\end{equation}

\begin{remark}
% Intuitively, the injected input will cause extra deviation in the state evolution of the NS, and the judgement criterion \eqref{eq:input_time} leverages the second-order information of this process to detect the injection. 
% This idea is very similar to the edge detection techniques in image processing community. 
% The reason of using xxxxx is as follows. 
Intuitively, the judgement criterion \eqref{eq:input_time} leverages the second-order information of the state evolution in the dynamic process. 
% this process to detect the injection. 
When there are no inputs involved, the states of all nodes will converge smoothly in an  exponential rate, and the differences $\Delta y_{t}^{i}$ and $\Delta y_{t-1}^{i}$ are very close to each other. 
When the input is injected, it will cause extra state deviation, which is reflected in $\Delta y_{t}^{i} - \Delta y_{t-1}^{i}$. 
We observe that the designed criterion is more sensitive to detect the input injection than the following detection criterion in \cite{qing2024topology}
\begin{align}\label{eq:criterion00}
|\Delta y_{t}^{i} |\ge \max\{ | y_{t\!-\!1}^{\ell } - y_{t-1}^{j}|\!: j,\ell \!\in\! \mathcal{V}\} + 3\sqrt{2}\sigma,
\end{align}
which essentially only utilizes the first-order information. 
This statement will also be verified in Sec. \ref{sec:simulation}. 
\end{remark}

Next, we demonstrate how to obtain an initial estimate about $W_{[i]}$ for all $i\in\mathcal{V}$. 
To begin with, this step is formulated as solving the following problem 
\begin{subequations}
\label{eq:initial_W}
\begin{align}
{\Phi}_{1_a}:~&\mathop {\min }\limits_{ W_{[i]}} ~  F_i(W_{[i]}) \\
&\text{s.t.}~~W_{[i]} \mathbf { 1} =1,
\end{align}
\end{subequations}
where $F_i(W_{[i]})$ represents the objective function associated with the available observations, and has different forms depending on the injection time $\tau_i$. 
% Recall that the set $\{ y_1,\cdots,y_{\tau_i} \}$ consists of the observations about node $i$ that are not affected by the latent input. 
If $\tau_i\ge n$, then we can directly model $F_i(W_{[i]})$ as the ordinary least squares function $\sum_{t = 1}^{\tau_i}( y_t^i - W_{[i]} y_{t-1} )^2$. 
% $\hat{W}_{[i]}^{(0)}$ by solving 
% \begin{subequations}
% \label{eq:initial_W}
% \begin{align}
% &\mathop {\min }\limits_{ W_{[i]}} ~  \sum\limits_{t = 1}^{\tau_i}( y_t^i - W_{[i]} y_{t-1} )^2 \\
% &\text{s.t.}~~W_{[i]} \mathbf { 1} =1. 
% \end{align}
% \end{subequations}
However, if $\tau_i<n$, the former objective function will lead to non-unique solutions. 
An effective way to deal with this issue is to introduce regularization terms along with the least squares. 
% Although it does not necessarily indicate that there are no inputs in the moments $\{t: t>\tau_i, ~ t\in\mathcal{T}_0^i  \}$, the magnitudes of these potential inputs are small and will not influence the state evolution significantly (i.e., \eqref{eq:s_bound} is not violated). 
% Hence, when $\tau_i<n-1$, this part of observations can be leveraged for the inference. 
Based on the above two points, the objective function $F_i(W_{[i]})$ is explicitly written as 
\begin{equation}
F_i(W_{[i]})\!=\!\left\{
\begin{aligned}
&\sum\limits_{t = 1}^{\tau_i}( y_t^i \!-\! W_{[i]} y_{t-1} )^2,&&\!\!\!\!\text{if}~\tau_i\!\ge\! n \\
&\sum\limits_{t = 1}^{\tau_i} ( y_t^i \!-\! W_{[i]} y_{t-1} )^2 + \!\rho \| W_{[i]} \|_1,&&\!\!\!\!\text{if}~\tau_i\!<\! n 
\end{aligned}\right. ,
\end{equation}
where the sparsity is motivated to avoid multiple solution cases in the second row, and is sufficient to yield a rough initial estimate. 
It is worth noting that there is rare possibility that $\tau_i$ is inferred to be empty by using \eqref{eq:input_time}. 
In this situation, we turn to set $\tau_i=n$ for the initial estimate. 

% \begin{remark}
% It should be noted that 
% \end{remark}

\subsection{Inference Algorithm by Using Zero-input Response}

In this part, we demonstrate how to obtain the final topology by inferring the inputs and using zero-input response. 

% To make full use of the valuable information of all the observations, an iterative inference strategy that estimates the unknown input is proposed. 

% version 1: 

First, let $\tilde{W}_{[i]}$ be the solution to the problem \eqref{eq:initial_W}, and the unknown input on node $i$ at time $t$ is calculated by
\begin{equation}\label{eq:input_compute}
\tilde{u}_t^i = 
\left\{
\begin{aligned}
&y_{t+1}^i-\tilde{W}_{[i]} y_{t}, &&\text{if}~t\ge \tau_i \\
&0, &&\text{otherwise}
\end{aligned}\right..  
\end{equation}
Note that the input series $\{\tilde{u}_t^i\}_{t=0}^{T-1}$ are directly polluted by observation noises, and do not well reflect the smooth characteristic of the input. 
To further mitigate this issue, we fit the input series $\{\tilde{u}_t^i\}_{t=0}^{T-1}$ by a series of polynomial basis functions, where the associated coefficients of the basis functions need to be estimated. 
Let $\bm{\theta}_i \in \mathbb{R}^{p+1}$ and $\bm{f}(t)\in\mathbb{R}^{p+1} $ ($p$ is the highest order of the polynomial) be the coefficient and basic function vectors, respectively. 
% approximate the coefficients of $u_t^i$, which is represented by a series of basis functions. 
Many estimation algorithms (e.g., the least square estimate \cite{ding2005hierarchical})
 are computationally efficient to calculate $\bm{\theta}_i$ by solving
\begin{equation} \label{eq:coff}
\hat {\bm{\theta}}_i= \arg \mathop{\min} \limits_{ \bm{\theta}_i }\sum\limits_{t = \tau_i}^{T-1} { (\bm{\theta}_i^\mathsf{T} \bm{f}(t) -  \tilde{u}_t^i )^2  }.
\end{equation}
Based on the obtained coefficients, the fitted inputs for further inference are given by 
\begin{equation} \label{eq:final_input}
\hat{u}_t^i  =
\left\{
\begin{aligned}
&\hat {\bm{\theta}}_i^\mathsf{T} \bm{f}(t) , &&\text{if}~t\ge \tau_i \\
&0, &&\text{otherwise}
\end{aligned}\right..
\end{equation}

% \begin{equation}
% \hat{u}_t^i(k) = y_{t+1}^i-\hat{W}_{[i]}(k) y_{t},
% \end{equation}
After the fitted inputs on all nodes are collected, we turn to calculate the state estimates corresponding to the zero-state response, i.e., the second term in \eqref{eq:separation}. 
To facilitate the expression, a recursive function $g_t$ is defined as 
\begin{equation}\label{eq:gt}
g_{t}=\tilde{W}_0 g_{t-1}+ \hat{u}_{t-1},~\forall t\in\mathbb{N}^+,
\end{equation}
where $g_0(k)=\hat{u}_0(k)$ and $\tilde{W}_0=[\tilde{W}_{[1]}^\mathsf{T},\cdots,\tilde{W}_{[n]}^\mathsf{T} ]^\mathsf{T}$. 
Then, the state estimates corresponding to the zero-input response can be calculated by 
\begin{equation}\label{eq:zt}
z_t=y_t-g_{t}. 
\end{equation}

\begin{algorithm}[t]
 % \small
    \caption{ The zero-input response based topology inference algorithm}
    \label{algo-1}
    \begin{algorithmic}[1]
    \REQUIRE { Observations $\{y_t\}_{t=0}^{T}$, polynomial order $p$}
    \ENSURE {The topology matrix $\hat{W}$}
    \STATE Compute $\Delta y_{0}=y_{1}-y_{0}$, set $\tau_i=n$ for all $i\in\mathcal{V}$ ;
    \FOR {$t=1,\cdots,T-1$}
    {
        \STATE Compute $\Delta y_{t}=y_{t+1}-y_{t}$;
        \FOR { $i\in\mathcal{V}$ }
        {
            \IF {$ |\Delta y_{t}^{i} - \Delta y_{t-1}^{i} | \ge s_t$}
            {
                \STATE Set $\tau_i=t$, and break this loop;
            }
            \ENDIF
        }
        \ENDFOR
    }
    \ENDFOR
    \FOR {$i\in\mathcal{V}$}
    {
        \STATE Obtain $\tilde{W}_{[i]}$ by solving problem ${\Phi}_{1_a}$;
        \STATE Compute the temporary inputs $\{\tilde{u}_t^i\}_{t=0}^{T-1}$ by \eqref{eq:input_compute};
        \STATE Obtain the final inputs $\{\hat{u}_t^i\}_{t=0}^{T-1}$ by polynomial approximation \eqref{eq:coff} and \eqref{eq:final_input};
    }
    \ENDFOR
    \STATE Use $\{\hat{u}_t\}_{t=0}^{T-1}$ to compute the zero-input response $\{z_t\}_{t=0}^{T}$ by \eqref{eq:gt}-\eqref{eq:zt};
    \STATE Obtain $\hat{W}$ by solving problem ${\Phi}_{1_b}$ from $\{z_t\}_{t=0}^{T}$. 
    \end{algorithmic}
\end{algorithm}

Next, with all the states $\{z_t\}_{t=0}^T$ collected, we design a decreasing weight based method to obtain the final topology. 
% Supposing that the zero-input response $\{z_t\}_{t=0}^T$ is estimated accurately, 
Based on the zero-input response, the underlying topology matrix is desired to satisfy the equation $z_{t} =W z_{t-1}$, which can be formulated as minimizing the least squares $\sum_{t=1}^{T}\|z_{t} -W z_{t-1} \|_2^2$ (it is also called as the empirical risk minimization). 
It should be pointed out that different weight settings on the terms $\{z_{t} -W z_{t-1}\}_{t=1}^T$ indicate a trade-off among different observations and yield different inference results. 
Here we adopt a decreasing-weight setting in the empirical risk minimization, given by
\begin{subequations}\label{eq:final_problem}
\begin{align}
{\Phi}_{1_b}:~&\mathop {\min }\limits_{W} ~\sum\limits_{t = 1}^{T}{ \alpha_{t} \left\|z_{t }\!-\!W z_{t-1} \right\|_2^2  }    \\ 
&\text{s.t.}~ W \mathbf{1} = \mathbf{ 1},
\end{align}
\end{subequations}
where the weights $\{\alpha_{t}\}_{t=1}^T$ satisfy $1\ge \alpha_1 >\cdots>\alpha_T>0$.

\begin{remark}
The reasonability of the decreasing weight design lies in two aspects. 
On the one hand, it is proved that in the ideal zero-input response, the difference $\|z_{t}-z_{t-1}\|_2$ will decay to zero as $t\to\infty$, which indicates the larger $t$ is, the less informative the term $\|z_{t}-z_{t-1}\|_2^2$ will be for the topology inference. 
On the other hand, since $z_t$ is calculated by filtering the accumulated inputs in the original observation $y_t$, the input estimation error may accumulate and be transmitted into $z_t$. 
This point also motivates us to assign lower weights on term $\|z_{t}-z_{t-1}\|_2^2$ with larger $t$. 
\end{remark}

Finally, the whole procedures to infer the topology are summarized in Algorithm \ref{algo-1}. 
Notice that numerous settings for the weights $\{\alpha_t\}_{t=1}^{T}$ are available to solve the problem \eqref{eq:final_problem}, and here we take the form $\alpha_t=1/t$ in the algorithm. 
% Specifically, the weights $\{\alpha_t\}_{t=1}^{T}$ are set as $\alpha_t=1/t$. 

% version 2: 

% The second step is to approximate the coefficients of $u_t^i$, which is represented by a series of basis functions. 
% Many estimation algorithms (e.g., the least square estimate \cite{ding2005hierarchical})
%  are computationally efficient to calculate $\bm{\theta}_i$ by solving
% \begin{equation} \label{eq:coff}
% \hat {\bm{\theta}}_i(k)= \arg \mathop{\min} \limits_{ \bm{\theta}_i(k) }\sum\limits_{t = \tau_i}^{T} { (\bm{\theta}_i^\mathsf{T}(k) \bm{f}(t) -  \tilde{u}_t^i )^2  }.
% \end{equation}
% Based on the fitted coefficient, the state with input filtering at $k$-th iteration is given by 
% \begin{equation}\label{eq:new_z}
% z^i_t(k)= \begin{cases}
% y^i_t(k),&\text{if}~t \le \tau_i,\\
% y^i_t(k)- \hat{\bm{\theta}}_i^\mathsf{T}(k) \bm{f}(t-1)  ,&\text{if}~t\ge \tau_i .
% \end{cases}
% \end{equation}
% Finally, by collecting the filtered states $\{z_t(k)\}_{t=0}^T$ and observations $\{y_t\}_{t=0}^T$, the topology matrix is estimated based on the following ideal equality
% \begin{equation}
% Z(k)=WY,
% \end{equation}
% where $Z(k)=\{z_1(k),z_2,\cdots,z_T \}$. 
% This case resembles the case in the last section and can be solved by the proposed two-layer optimization. 

\begin{figure}[t]
\begin{center}
\includegraphics[width=0.4\textwidth]{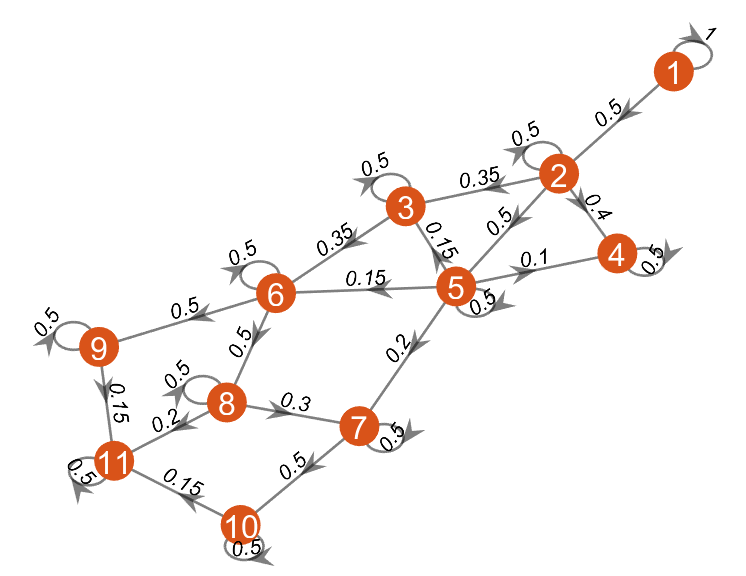}
\end{center}
\vspace{-8pt}
\caption{The topology of the NS of $11$ nodes. The directed edges are drawn in gray color along with their weight values.}
\label{fig:topo_setting}
\vspace{-12pt}
\end{figure}

\begin{figure}[t]
\begin{center}
\includegraphics[width=0.41\textwidth]{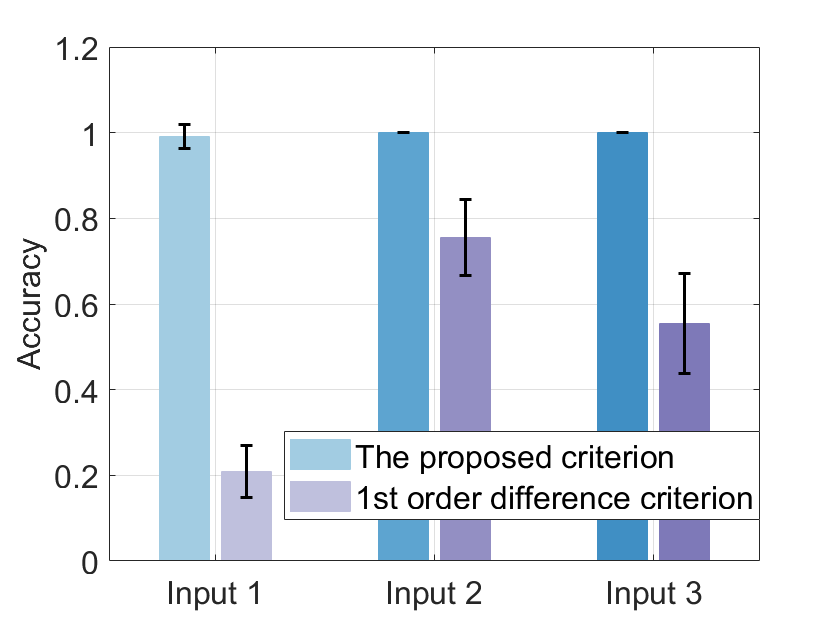}
\end{center}
\vspace{-8pt}
\caption{The accuracy comparison of injected input detection criteria.}
\label{fig:input_detection}
\vspace{-12pt}
\end{figure}

\begin{figure*}[t]
\centering
\subfigure[Under input type-I]{\label{fig:input1}
\includegraphics[width=0.33\textwidth]{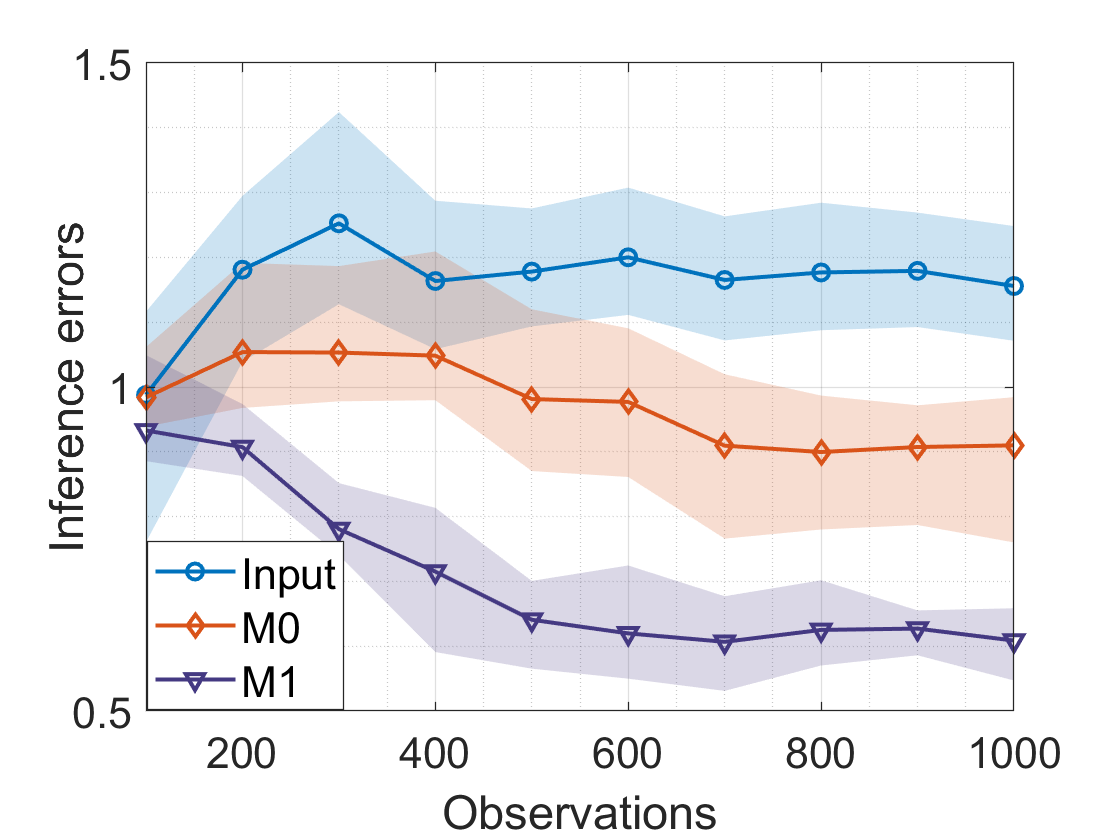}}
\hspace{-9pt}
\subfigure[Under input type-II]{\label{fig:input2}
\includegraphics[width=0.33\textwidth]{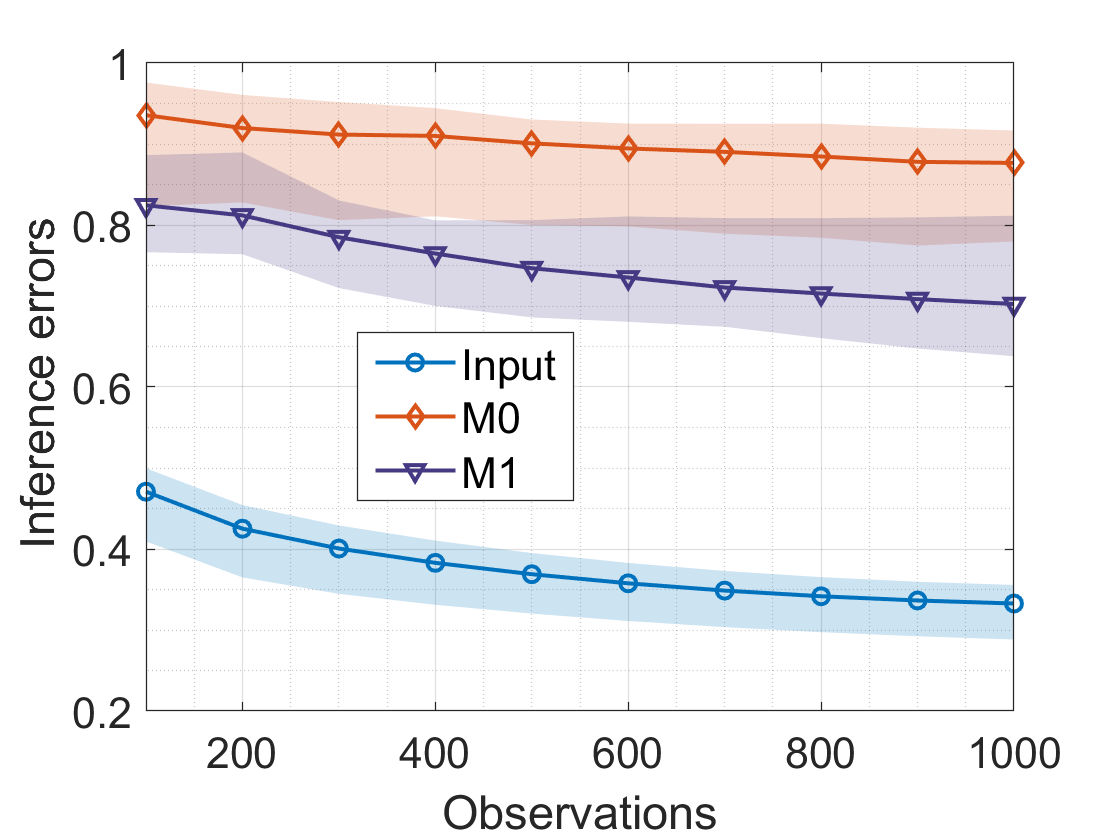}}
\hspace{-9pt}
\subfigure[Under input type-II]{\label{fig:input3}
\includegraphics[width=0.33\textwidth]{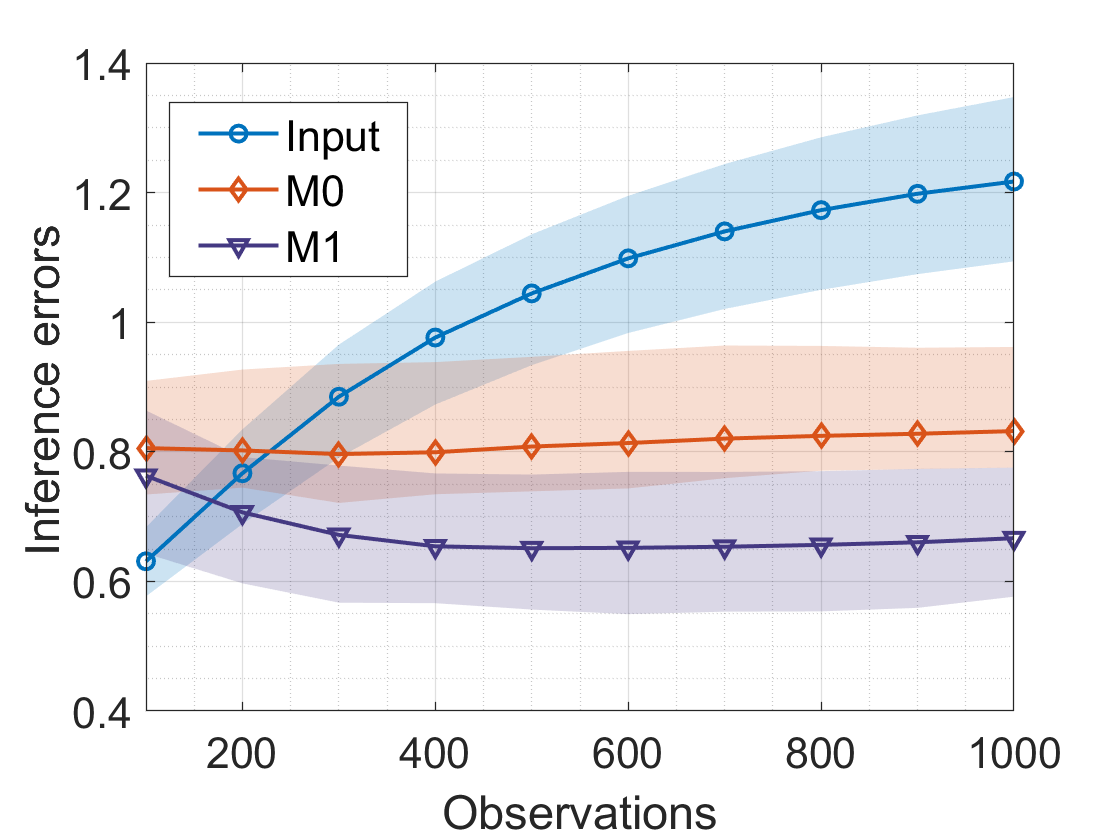}}
\caption{Topology inference performance under different input types.}
\label{fig:input_test}
\vspace*{-10pt}
\end{figure*}

\begin{figure*}[t]
\centering
\subfigure[NMSE]{\label{fig:NMSE}
\includegraphics[width=0.33\textwidth]{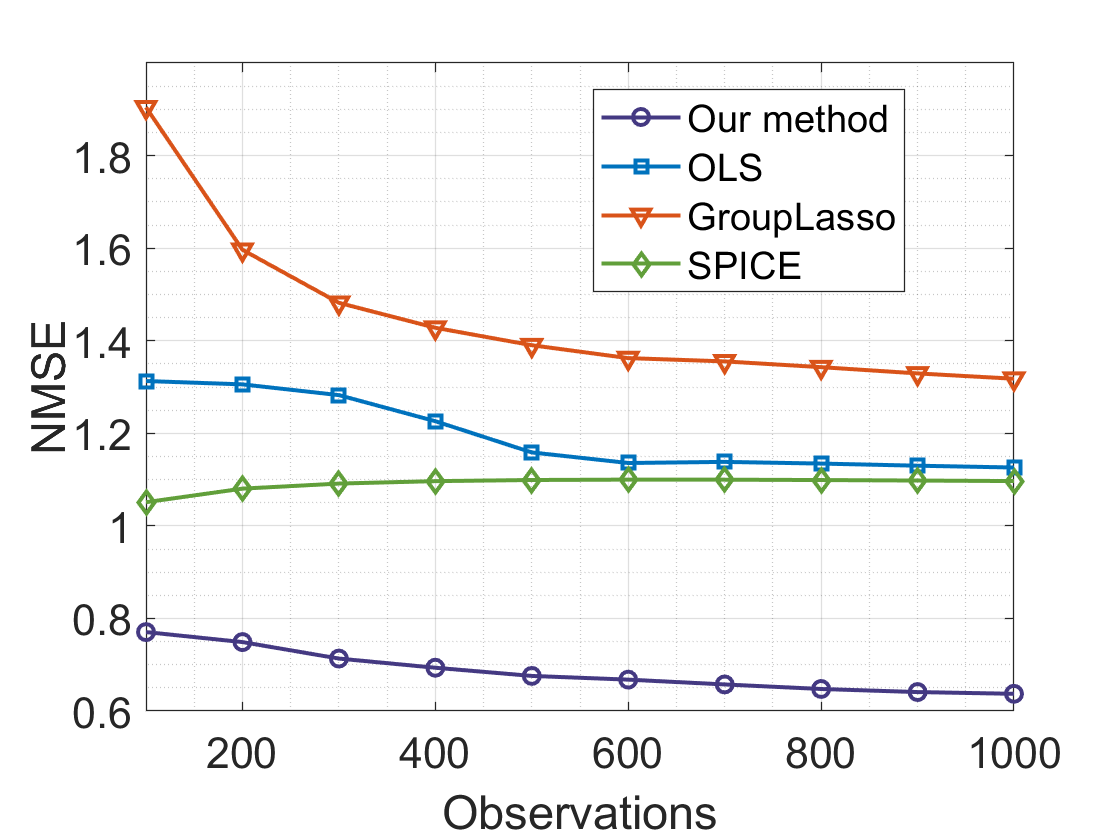}}
\hspace{-9pt}
\subfigure[EIER]{\label{fig:EIER}
\includegraphics[width=0.33\textwidth]{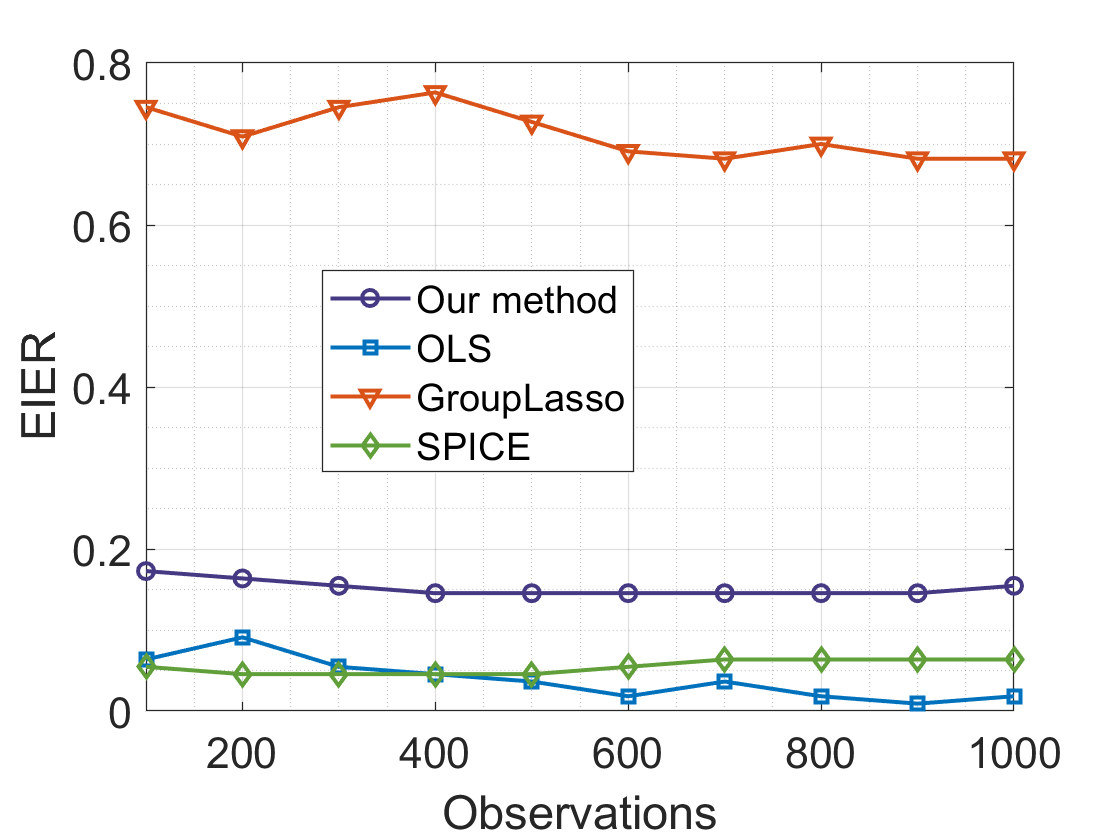}}
\hspace{-9pt}
\subfigure[F-score]{\label{fig:score}
\includegraphics[width=0.33\textwidth]{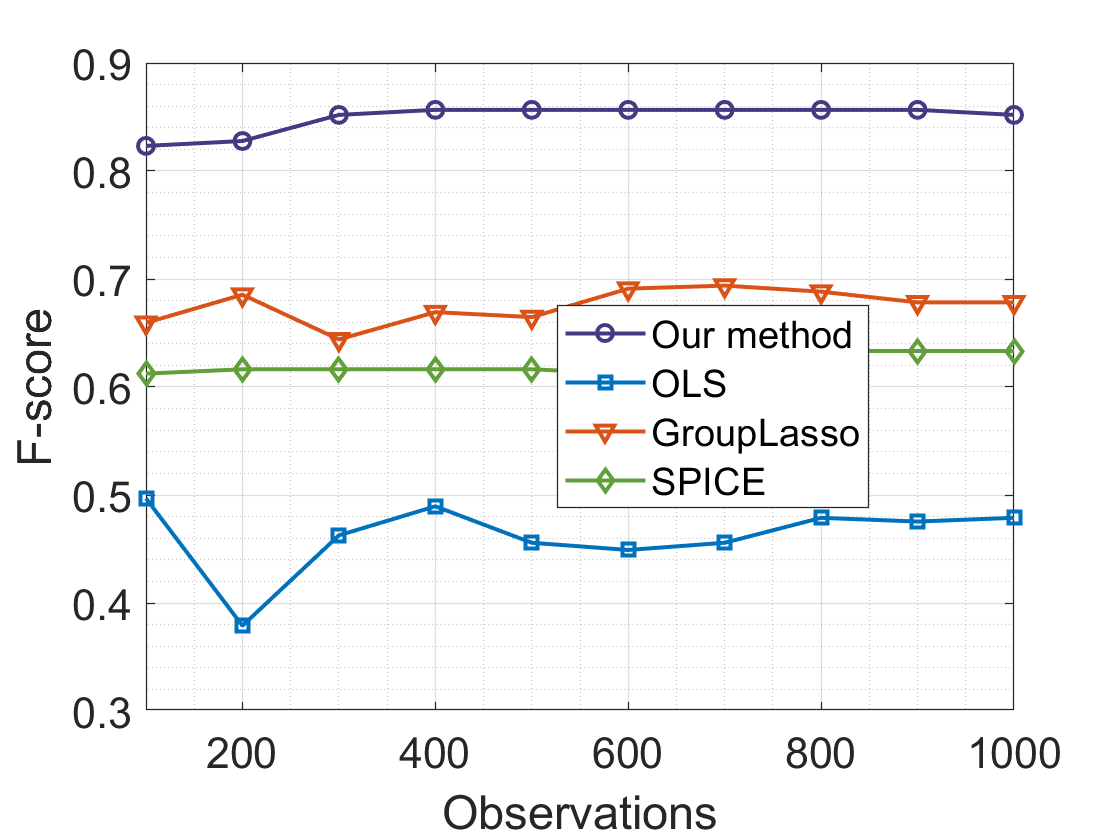}}
\caption{Topology inference performance comparisons using different methods.}
\label{fig:comparison}
\vspace*{-10pt}
\end{figure*}

\section{Simulations}\label{sec:simulation}

In this section, extensive simulations are conducted to illustrate the effectiveness of the proposed method. 

\textit{Simulation setting.} 
We use a directed network of $n=11$ nodes, whose topology structure is drawn in Fig.~\ref{fig:topo_setting}. 
The initial states of the NS are evenly spaced from $100$ to $600$, and the variance of the observation noise is set as $\sigma^2=0.5^2$. 
As for the inputs for $i\in\mathcal{V}$, three kinds of time-varying functions are considered, given by
\begin{align}
u_{t,1}^i &=a_i \cos ({2\pi t b_i}/{T}), \nonumber \\
u_{t,2}^i &=a_i \exp\left(-b_i t \right) +c_i, \nonumber \\
u_{t,3}^i &=a_i \exp\left(-b_i t\right) + c_i t +d_i, \nonumber 
\end{align}
where the coefficients $a_i$, $b_i$, $c_i$, and $d_i$ are randomly generated from intervals $[10,20]$, $[1,10]$, $[-5,5]$ and $[-5,5]$, respectively. 
The input is injected into the NS when $t=7$, which is smaller than $n$.

\textit{Results analysis}. 
First, we compare the proposed input detection criterion \eqref{eq:input_time} with the first-order difference criterion \eqref{eq:criterion00}. 
The ratio of the correct input detection number to the overall test number is used to describe the accuracy. 
After running the test $20$ times from the same initial states for each input setting, the detection accuracy of the two criteria are drawn in Fig.~\ref{fig:input_detection}, where the height of a colorful bar represents the mean detection rate and the black `I' line represents the standard deviation range around the mean point. 
It is straightforward to find that the proposed criterion is more effective in input injection detection and achieves much higher accuracy in all three input settings.

Next, to demonstrate the benefits of using estimated zero-input response (denoted as \textbf{M1}), the following method (denoted as \textbf{M0}), which directly subtracts the one-moment input in each step, is simulated as a comparison
\begin{subequations}
\begin{align}
\textbf{M0}:~&\mathop {\min }\limits_{ W_{[i]} } ~\sum\limits_{t = 1}^{T } \left\| \tilde{z}_{t}^i - W_{[i]} \tilde{z}_{t-1} \right\|_2^2 \\
&\text{s.t.}~~W_{[i]} \mathbf { 1} =1, 
\end{align}
\end{subequations}
where the element of the state variable $\tilde{z}^i_t$ is calculated by 
\begin{equation}\label{eq:new_z}
\tilde{z}^i_t= \begin{cases}
y^i_t,&\text{if}~t \le \tau_i\\
y^i_t- \hat{\bm{\theta}}_i^\mathsf{T} \bm{f}(t-1)  ,&\text{if}~t> \tau_i 
\end{cases}.
\end{equation}
Then, Fig.~\ref{fig:input_test} shows the inference errors of the two methods under three kinds of inputs. 
The experiments are conducted $20$ times using the same initial states, and the color bar and curve represent the error interval and mean, respectively. 
Note that the two methods utilize the same estimated inputs, whose errors are calculated by $\sum_{t=0}^{T-1}\|\hat{u}_t-u_t\|_2^2/\sum_{t=0}^{T-1}\|u_t\|_2^2 $ and represented by the blue bar. 
The topology inference error is evaluated by the normalized mean square error (NMSE)
\begin{equation*}
\operatorname{NMSE}=\|\hat{W}-W\|_F/\|W\|_F. 
\end{equation*}
The plotted curves demonstrate that in all three input settings, the proposed method outperforms \textbf{M0} that only filters the input once in each step, corroborating the effectiveness of using decreasing-weight design on the zero-input response.
% It is clear to see that the proposed method outperforms the \textbf{M0} in all tested input forms, which verifies the effectiveness of using decreasing-weight and zero-input states. 

Finally, we compare the proposed method with the popular ordinary least squares (OLS) method, GroupLasso algorithm in \cite{bolstad2011causal}, and SPICE algorithm in \cite{venkitaraman2019learning}. 
In this experiment, apart from the NMSE metric, we also use the edge identification error (EIER) and F-score for performance evaluation, calculated by
\begin{align}
&\operatorname{EIER}(\widehat{W}, W)={\|W-\widehat{W}\|_{0}}/({n(n-1)}), \nonumber \\
&\operatorname{FS}(\widehat{W}, W)={2 \mathrm{tp}}/({2 \mathrm{tp}+\mathrm{fn}+\mathrm{fp}}), \nonumber
\end{align}
where $\mathrm{tp}$, $\mathrm{fn}$, and $\mathrm{fp}$ denote the true positive number (i.e., the case
when the method detects an actual edge), the false negative number, and the false positive number, respectively. 
Note that F-score is commonly used in the binary classification scenario, and its value locates in $[0,1]$, where $1$ indicates perfect classification. 
Fig.~\ref{fig:comparison} presents the comparison results of the four methods, where the curves represent the mean of $20$ times of test using the same topology and input setting in the last example. 
It is clear to see that the proposed method achieves better inference performance in terms of the NMSE and F-score than those of other three, while the EIER performance is modest. 
We also observe that the performance curves are likewise when we use other input parameters and randomly-generated topologies satisfying Assumption \ref{assu:topo}, which validate the applicability of the proposed method.

\section{Conclusion} \label{sec:Conclusion}
In this paper, we investigated the topology inference problem of NSs when unknown time-varying inputs are injected into the network. 
First, we designed a detection criterion with probability guarantees to identify the input injection time.  
Then, we used the determined injection time to obtain an initial topology estimate, and proposed a recursive input filtering algorithm to further approximated zero-input response. 
Finally, we formulated a decreasing-weight based optimization problem to infer the topology from zero-input response. 
We further provided comprehensive simulations to illustrate the effectiveness of the proposed method.

One of the promising future directions is to infer the topology by utilizing the state responses to the unknown inputs, rather than filtering the inputs from observations. 
The challenges mainly lie in how to discriminate the zero-state responses and further enhance the inference accuracy.

% In this paper, we proposed TO-TIA and IE-TIA to learn the directed network topology from the observations of cooperation dynamics, where the latent input injected into the network is unmeasurable.
% We first proposed TO-TIA towards the time-invariant input:
% i) an input filtering method was designed to eliminate the influence of latent input on the dynamic process, and
% ii) a two-layer optimization strategy was proposed to improve the topology inference accuracy.
% We then proposed IE-TIA towards the time-varying input:
% i) an identification method was designed to identify all the injected agents,
% and ii) an iterative estimation strategy was proposed to improve the accuracy of the estimation of the network topology and latent input.
% Furthermore, we provided strict theoretical analysis and comprehensive simulations to illustrate the effectiveness and the accuracy of TO-TIA and IE-TIA.

% One of the promising future directions of this work is to utilize the latent input to steer the network to a target state.
% The potential challenges mainly lie in how to select the injected agents and guarantee the controllability of the dynamical system.
% We believe that by paying off efforts, new favorable alternatives will be provided for multi-agent cooperative control.

%% 是否需要notation，示意图，算法框图，notation表

\bibliographystyle{ieeetr}

\end{document}